\documentclass[a4paper,showpacs,preprintnumbers,floatfix,twocolumn,prb,longbibliography]{revtex4-1}

\usepackage{miller}
\usepackage{epsfig}
\usepackage{color}
\usepackage{float}
\usepackage{graphicx}
\usepackage[utf8]{inputenc}

\definecolor{red}{rgb}{1.0,0.0,0.0}
\definecolor{blue}{rgb}{0.0,0.0,1.0}

\newcommand{\ang}[1]{#1$^{\circ}$}

\begin{document}

\title{Formation of parallel and perpendicular ripples on solid amorphous surfaces by ion beam-driven atomic flow on and under the surface}

\author{Alvaro Lopez-Cazalilla} \email{alvaro.lopezcazalilla@helsinki.fi}
\affiliation{Helsinki Institute of Physics}
\affiliation{Department of Physics, P.O. Box 43, FI-00014 University of Helsinki, Finland}
\author{Kai Nordlund}
\affiliation{Department of Physics, P.O. Box 43, FI-00014 University of Helsinki, Finland}
\author{Flyura Djurabekova}
\affiliation{Helsinki Institute of Physics}
\affiliation{Department of Physics, P.O. Box 43, FI-00014 University of Helsinki, Finland}

\date{\today}


\begin{abstract}
{The off-normal ion irradiation of semiconductor materials is seen to induce nanopatterning effects. Different theories are proposed to explain the mechanisms that drive self-reorganization of amorphisable surfaces. One of the prominent hypothesis associates formation of nanopatterning with the changes of sputtering characteristics caused by changes in surface morphology. At ultra-low energy, when sputtering is negligible, the Si surface has still been seen to re-organize forming surface ripples with the wave vector either aligned with the ion beam direction or perpendicular to it.} 
{In this work, we investigate the formation of ripples using molecular dynamics in all the three regimes of ripple formation: low angles where no ripples form, intermediate regime where the ripple wave vectors are parallel to the beam, and high angles where they are perpendicular to it. We obtain atom-level insight on how the ion-beam driven atomic dynamics at the surface contributes to organization, or lack of it, in all the different regimes.} Results of our simulations agree well with experimental observations in the same range of ultra-low energy of ion irradiation.

\end{abstract}

\maketitle

\section{Introduction}
\label{sec:intro}

Surface self-organization is an interesting phenomenon and seen at different scales in time and in space, if the surface is exposed to momentum transfer at tilted incidence. 
For instance, the blowing wind or the running water can develop ripples on sand surfaces the ripples 
aligned either parallel or perpendicular to the wind direction \cite{Fou10}. The mechanisms behind this was explained by Bagnold \cite{bagnold1941}, observing that the grains move according to the \textit{reptation} and \textit{saltation} processes. 

Under ion beam irradiation of semiconductor materials, the momentum is transferred to the surface via individual ion impacts. Collectively, these impacts may induce patterning on the surface of the material similar to the sand ripples and dunes. The ion-beam induced nanopatterning effect was observed the first time already in the 1960s by Cunningham \cite{Cun60} in Au using 8 keV-Ar$^{+}$ at \ang{70} off-normal incidence. Since then, many theories have been developed in order to explain the ripple formation with the focus on erosion \cite{bradleyharper1988}, atom redistribution \cite{dc4,norrisnature} and stress build-up \cite{norris2012a,norris2012b,morenobarrado2015,munozgarcia2019}. The structures observed in experiments range from ordered nanodots in GaSb \cite{fackso1999} and in Si \cite{DebasreeSinanodots} (with sample rotation) to parallel \cite{hofsass2013vel} and perpendicular to the ion-beam projection ripples {(i.e. the wave vector direction)} \cite{Lop18}. 

Computational methods like Molecular Dynamics (MD) or Binary Collision Approximation (BCA) have been used extensively to study this curious phenomenon. The application of these methods has shed light on some of the aspects of it. Norris et al. \cite{norris2009} introduced the crater function formalism which allowed to predict the pattern-formation and corresponding wavelength of these periodic structures \cite{norrisnature,Lop17,Lop18}. This model was particularly successful if it was based on single-impact MD simulations, emphasizing the importance of small atomic displacements which result from low-energy multi-body interactions accessible in MD, but not in BCA \cite{Lop17}. 

Although single impact simulations have already provided important insights in understanding the mechanisms of nanopattern formation, the main advantage of the MD method is a possibility to simulate cumulative radiation effects due to high fluence irradiation. Under the time and length scale limitation of this type of atomistic modelling, it is, nevertheless, possible to simulate sequential ion impacts to understand the relative contribution of erosion and atomic redistribution on surface modification during prolonged ion irradiation with different energies and at different incident angles \cite{Lop18}. At the same time, we measure the relative effect of structural order of irradiated material with respect to efficiency of surface modification 
\cite{Lop18b}.  
The results show that the contribution of the erosion contrary to redistribution, is strongly dependent on the material and the ion energy. 
Recently it was shown that very high fluences in MD simulations can be achieved by using a speed-up algorithm \cite{Fri16}. 
Applying this algorithm we were able to observe directly the pattern formation in parallel mode on a-Si surface under very low incident energies of 30 eV \cite{Lop18a} similar to the ripples observed experimentally \cite{Lop18}. 
{However, the reason why no ripples form at low incidence angles and the formation mechanism of perpendicular mode ripples (i.e. ripples oriented in the same direction as the ion beam) under very small grazing incidence remained still unclear.}

\begin{figure}[H]
\begin{center} 
\includegraphics[width=1\linewidth]{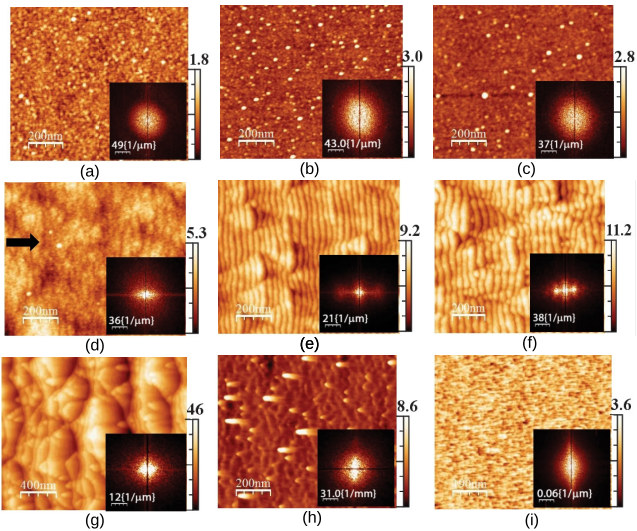}
\end{center}
\caption{\label{fig:exp_30eV_Ar_Si}
AFM images of the evolution of morphology at different ion incidence angles: (a) 0$^{\circ}$, (b) 25$^{\circ}$, (c) 35$^{\circ}$, (d) 55$^{\circ}$, (e) 65$^{\circ}$, (f) \ang{70}, (g) 75$^{\circ}$, (h) 80$^{\circ}$ and (i) \ang{85} for 30 eV-Ar$^+$ ion irradiation, current density 52 $\mu A$ cm$^{-2}$ and fluence 1$\times$10$^{19}$ ions cm$^{-2}$. The different insets show the corresponding FFT images and the black arrow indicates the beam direction for oblique incidence ion irradiation. The ripples in images (e) and (f) are called parallel mode ripples because the wave vector (see insets) is parallel to the ion beam. The ripples in image (i) are correspondingly called perpendicular mode ripples. {Reprint from Ref. \citenum{Lop18}.}}
\end{figure}


In this work, we examine systematically ripple formation over all incident angles to understand why there are no ripples at low incidence angles, and why the ripples turn direction at the highest angles. To understand the formation mechanisms comprehensively, we analyze the atom redistributions separating displacements in the three different dimensions, and in particular examine in detail displacements at very small grazing angles (the incident angle $>$ \ang{80}), where the ripples in the perpendicular mode are expected to appear~\cite{madi2012}. Moreover, to enable a quantitative analysis of whether self-organization has occurred, we also perform perform Fast Fourier Transform (FFT) of the surfaces.



Different quantitative analyses of the simulated data along with the direct observation of 
the evolution of the irradiated surfaces 
complement the model presented in \cite{Lop18a} and allow us to complete the picture of ripple formation on amorphous surfaces within a certain range of incident angles at least, at very low near the displacement threshold energies.

\section{Methods}
\label{sec:methods}

A flat-surface cell of amorphous Si containing 73584 atoms was relaxed \cite{Lop18,Lop18b} to 300 K in order to perform the sequential bombardment. The final size of the cell is $16.56\times16.56\times5.15$ nm$^3$.

The simulations were performed using the PARCAS MD code \cite{Gha97,Nor97c}. The Si-Si interactions are described using the environment-dependent inter-atomic potential (EDIP) \cite{justopot1997,justopot1998}, complemented at short distances by the purely repulsive ZBL potential \cite{nordarsi}. The same type of repulsive potential describes the Ar-Si interaction. The bombardment of the sample is done sequentially \cite{Lop18,Lop18b}, so we used the pair potential with high energy repulsive part from DFT DMol calculations \cite{nordarsi} smoothly joined to the LJ equilibrium part \cite{Kittel} at larger distances to describe the Ar-Ar interactions.

The irradiation of the sample is done using the speed-up scheme \cite{Fri16}, which allows for reaching high fluences using MD simulations \citep{Lop18a}. In this study, we used four angles of ion incidence, $\theta$= $\{0^{\circ}, 45^{\circ}, 70^{\circ} 85^{\circ}\}$. The normal incidence is used as a reference case to show how the surface behaves during ion irradiation up to high fluences when the ion beam is not tilted. $\theta=45^{\circ}$ provides sufficiently strong tilt, while yet being less than the threshold $\theta=55^{\circ}$ \cite{lozano2013}. {$\theta=70^{\circ}$ corresponds to the regime where parallel mode ripples are expected to form (the MD results at this angle were the same as those in \cite{Lop18a}, however in this work we performed additional displacement and Fourier transform analysis on the MD results).} $\theta=85^{\circ}$ corresponds to the very small grazing incidents, at which the ion is still able to hit the surface. At this angle, the perpendicular mode of ripples was observed in experiments \cite{Lop18} shown in Figure \ref{fig:exp_30eV_Ar_Si} (i). 

The 30 eV-Ar$^+$ atom is placed above the surface at such a location that when it hits the surface the impact is always at the center independently of the incident angle. The center of the surface is selected for the impact to initiate the cascade far from the border to avoid the interactions over periodic boundaries. For consistency with the tilt of the beam, the azimuthal angle is always fixed and is aligned with the $x$ axis. Despite the fixed entry point, the ion actually impacts always at a random place of the cell because of shifting the cell in both $x$ and $y$ directions prior to every impact. 

We simulated 54000 consecutive ion impacts at the 0$^{\circ}$, 45$^{\circ}$ and 70$^{\circ}$ reaching the fluence $2\times10^{16}$ ions cm$^{-2}$, which was sufficient to observe the high-fluence effects. In the case of \ang{85}, the grazing angle is too small and much higher fluence was needed to observe self-organization of irradiated surface. In this case, we simulated 160000 consecutive Ar impacts (fluence of $5.9\times10^{16}$ ions cm$^{-2}$). This is not surprising; in experiments, self-organization of the irradiated surface with grazing angle was also observed at much higher fluences than those for the incident angles closer to the normal \cite{Lop18}.

After every nine impacts, the temperature of the system is restored to 300 K applying the Berendsen thermostat \cite{berendsen} during additional 30 ps. Each individual impact is allowed to develop for 1 ps only with the thermostat controlling temperature to 300 K applied to the 0.8 nm-thick layers at the borders of the cell in the $x$ and $y$ directions, but not at the surface. More details on the speedup scheme can be found in \cite{Lop18a}.  

The erosion and atomic redistribution components of the ripple formation mechanism are computed by counting the number of sputtered atoms in the system, and the total displacement. The total displacement was calculated summing up separately each component of the individual displacement vectors of all atoms displaced in a cascade: 
\begin{equation}
	\delta_{w} = \sum\limits_{i=1}^{N_{\hbox{\tiny{displaced}}}} (w^{i}_{\hbox{\tiny{final}}} - w^{i}_{\hbox{\tiny{initial}}}),
\end{equation}\label{eq:delta}

where $N_{\hbox{\tiny{displaced}}}$ is the amount of atoms displaced within the cell, the indices  $final$ and $initial$ refer to the current and initial positions of the atoms, respectively, while $w=\{x,y,z\}$. Moreover, under high fluence ion irradiation the accumulation of stresses is expected due to numerous atomic displacements in multiple cascades. This quantity is also accessible in MD as we showed in Ref. \citenum{Lop18a}. We attempted to analyze the stress in the current simulations as well, but could not observe any clear relation between the ripple formation and the stress build-up at this low energy.

\subsection{Fast Fourier Transform}\label{sec:FFT}


To obtain a diffraction-like image of the surfaces (similar to those obtained from the AFM experiments, the Fast Fourier Transform (FFT) is calculated using the MATLAB toolbox \cite{MatlabOTB}. We considered the last frame simulated at different irradiation angles, where the surface features are the most prominent, enhancing the probability of obtaining a more pronounced result. However, the ridge-trough distance in some of the cases is not marked, so we needed to magnify this difference. In order to do that, we consider 100$\times$100 bins in the $x-y$ plane, containing a small amount of atoms per bin. Among those atoms, we consider the highest atom ($z$-coordinate) in the bin to amplify the differences between the bins that are separated each other, and consequently, detect more efficiently the possible pattern. Then, the color-scale is chosen independently for all the cases, in order to highlight the orientation of any formation. 

\section{Results}
\label{sec:results}

\subsection{Surface self-organization}
\label{sec:results_directobservation}

\begin{figure}[H]
\begin{center} 
\includegraphics[width=1\linewidth]{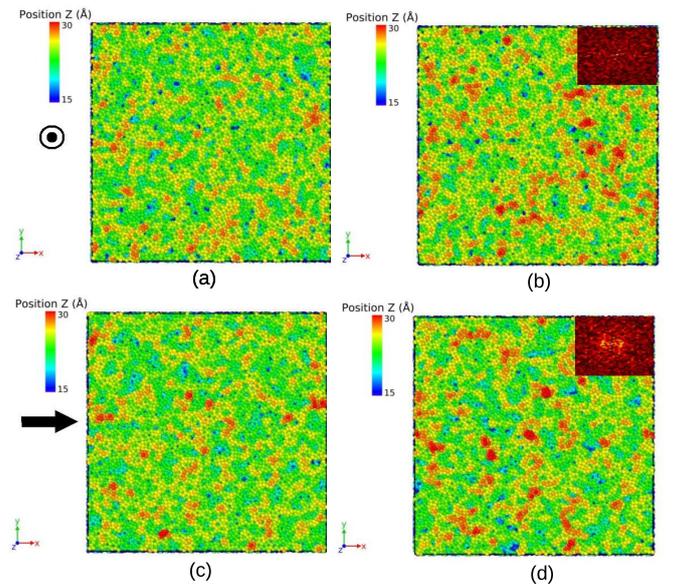}
\end{center}
\caption{\label{fig:pos_z_0d_45d}
Evolution of the a-Si surface under 30eV-Ar$^{+}$ at 0$^{\circ}$ (a) 4500 and (b) 54000 impacts (including FFT of the surface); and at \ang{45} (c) 4500 and (d) 54000 impacts (including Fast Fourier Transform (FFT) of the surface). The black arrow indicates the irradiation direction.}
\end{figure}

\begin{figure} 
\begin{center} 
\includegraphics[width=1\linewidth]{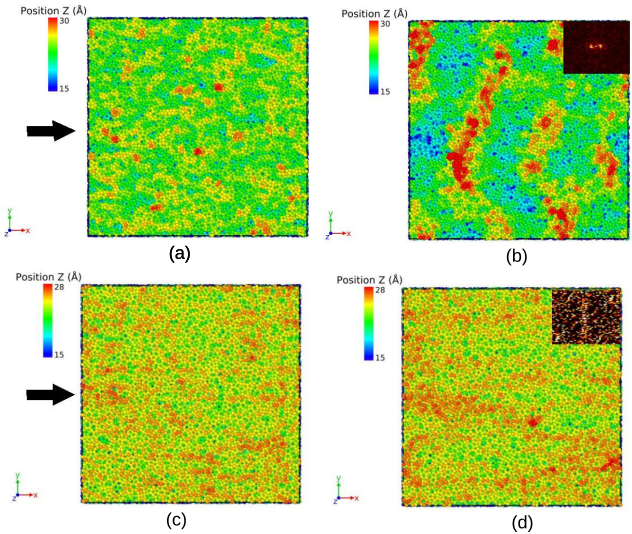}
\end{center}
\caption{\label{fig:pos_z_70d_85d}
Evolution of the a-Si surface under 30eV-Ar$^{+}$ at \ang{70} (a) 4500 and (b) 54000 impacts (adapted from result covered in Ref. \cite{Lop18a} and including FFT of the surface); and at \ang{85} (c) 78000 and (d) 160200 impacts (including FFT of the surface). The black arrow indicates the irradiation direction.}
\end{figure}

In Figures \ref{fig:pos_z_0d_45d} and \ref{fig:pos_z_70d_85d} we show the gradual evolution of a-Si surface under ion irradiation with different incidence. Figures \ref{fig:pos_z_0d_45d} and Figures \ref{fig:pos_z_70d_85d}(a,b) show the simulation snapshots after 4500 and 54000 ion impacts on the left and the right sides of the corresponding figures, respectively. Figures \ref{fig:pos_z_70d_85d}(c,d) show similar snapshots but after 78000 and 160200 ion impacts, respectively. In the top corner of Figures \ref{fig:pos_z_0d_45d}(b,d) and \ref{fig:pos_z_70d_85d}(b,d) we also show the FFT images generated from the analysis of the atoms in the corresponding surfaces for better comparison with the experimental results shown in Figure \ref{fig:exp_30eV_Ar_Si}.

First of all, we observe that under the normal incidence the surface roughness develops rapidly already at the early stage of the irradiation (see Figure \ref{fig:pos_z_0d_45d}(a)). With fluence increase, the roughness increases as well, forming randomly located mounds, similarly to those observed initially under the \ang{70} incidence in the previous simulations~\cite{Lop18a}. However, under the normal incidence the mounds do not self-organize, but remain random even at fluence more than an order of magnitude higher, see Figure \ref{fig:pos_z_0d_45d}(b). This behaviour persists until the end of the simulation. The absence of a clear surface pattern in this case is consistent with the AFM measurements in Figure \ref{fig:exp_30eV_Ar_Si}(a). The surface evolution under normal incidence can be seen in the supplementary movies: \textsc{0degrees-surf-formation-disp.mp4} and \textsc{0degrees-surf-formation-posz.mp4}.  
The random nature of roughness is also confirmed by the FFT image shown in the inset of Figure \ref{fig:pos_z_0d_45d}(b) with no preferential structure, which agrees well with the experimental FFT image in Figure \ref{fig:exp_30eV_Ar_Si}(a).  

After tilting the ion beam to \ang{45} off-normal, we also observe significant development of surface roughness at the early stages of irradiation (see Figure \ref{fig:pos_z_0d_45d}(c)). At later stage (Figure \ref{fig:pos_z_0d_45d}(d)), the random mounds start organizing in some surface features, however, the process does not complete in any  recognizable pattern even at the fluence of $\sim$ 2$\times$10$^{16}$cm$^{-2}$ (for comparison, the clear ripples were formed at the fluence 1.8$\times$10$^{16}$cm$^{-2}$ under the \ang{70} incidence as reported in Ref. \citep{Lop18a}). 
Local self-organized ridges seen in Figure  \ref{fig:pos_z_0d_45d}(d) do not align in any preferential direction, i.e. neither as parallel- nor perpendicular-mode ripples (see movies \textsc{45degrees-surf-formation-disp.mp4} and \textsc{45degrees-surf-formation-posz.mp4}). In the FFT image in the inset of Figure \ref{fig:pos_z_0d_45d}(d) we notice that some alignment is emerging. However, this alignment is rather vague, which is explained by the absence of clear periodicity. On the larger experimental surface, this vague pattern is expected to disappear due to higher randomness of orientations of local self-organized features.  Unfortunately, there is no experimental data at this angle in Figure \ref{fig:exp_30eV_Ar_Si}, although interpolation between the result shown in Figures \ref{fig:exp_30eV_Ar_Si}(c) and \ref{fig:exp_30eV_Ar_Si}(d) is consistent with the above argument. 

In Figure \ref{fig:pos_z_70d_85d}(a-b) we show how the ripples in parallel mode emerge from the initially random surface roughness at \ang{70} incidence. Here we also generated the FFT image for the ripples in Figure \ref{fig:pos_z_70d_85d}(b) that show a clear pattern similar to that obtained for the experimentally observed ripples, see inset of Figure \ref{fig:exp_30eV_Ar_Si}(f).  
Increasing the incidence angle further to \ang{85} off-normal slows down the dynamics of self-organization. In these simulations, we had to increase the fluence to much higher value to start observing any organizational pattern. Note the difference in the total numbers of ion impacts simulated for this case: Figure \ref{fig:pos_z_70d_85d}(c) shows the snapshot for the simulation cell after 78000 ion impacts (4500 ions for lower incident angles) and Figure \ref{fig:pos_z_70d_85d}(d) after 160200 ion impacts (54000 ions for lower incident angles). The necessity of the fluence increase is consistent with the experiments reported in Ref. \cite{Lop18} and is explained by much lower momentum transfer from ion beam to surface atoms, since many impacts result in back-scattering of incident ions. These presented snapshots showing the surface evolution under \ang{85} off-normal incidence, demonstrate the gradual self-organization of the surface into ripples in the perpendicular mode, although the roughening of the surface under the grazing incidence is much slower and shallower (see Figure \ref{fig:exp_30eV_Ar_Si} (i)). We observe formation of the first mounds at the beginning of the simulation (see \textsc{85degrees-surf-formation-posz.mp4}). 
As the fluence increases (Figure \ref{fig:pos_z_70d_85d} (c)), these mounds merge with others and some thin ripple-like structures start developing (the complete dynamics, see movies \textsc{85degrees-surf-formation-disp.mp4} and \textsc{85degrees-surf-formation-posz.mp4}). This surface reorganization becomes fully clear at the last stages of the simulation (Figure \ref{fig:pos_z_70d_85d} (d)), where the ridges of the ripples align along the projected ion beam forming ripples in the perpendicular mode i.e with the wave vector perpendicular to ion beam projection. This result is fully consistent with the experiment from Ref. \cite{Lop18} where the rotation of the ripples was also observed at \ang{85}. In addition to this, the FFT image provided in Figure \ref{fig:pos_z_70d_85d} (d)) is also showing certain perpendicular orientation, as the one shown in Figure \ref{fig:exp_30eV_Ar_Si} (i). 


\subsection{Erosive and redistributive components of surface self-organization}
\label{sec:results_analysis}


We further analyze quantitatively the effect of erosive and redistributive components to the ripple formation process in our simulations. 
For completeness of the comparative analysis, we also include the case of ripple formation under the \ang{70} incidence from Ref. \cite{Lop18a}. 


\begin{figure}[H]
\begin{center} 
\includegraphics[width=1\linewidth]{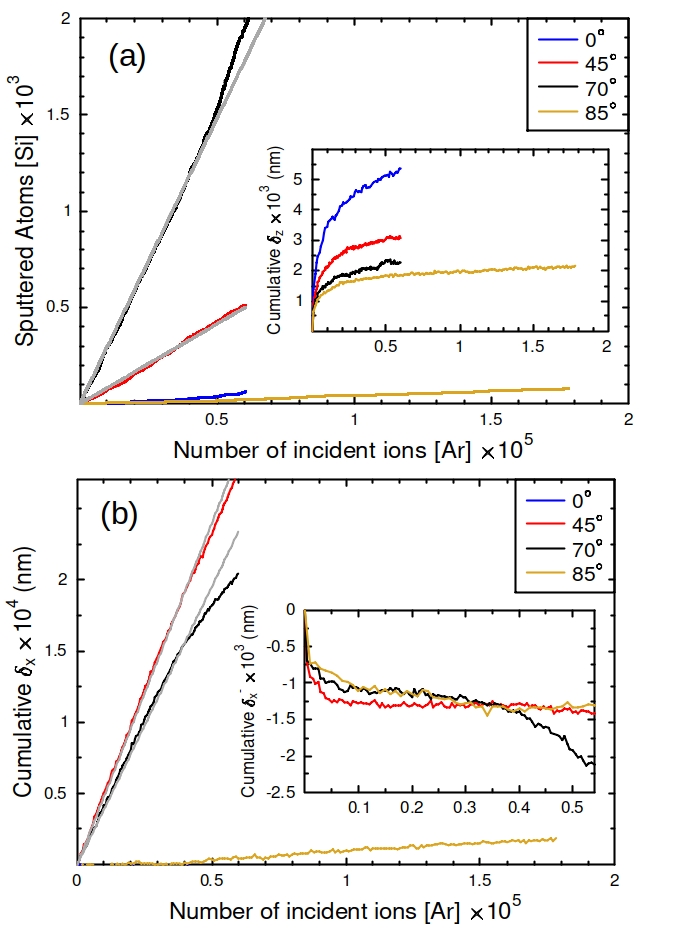}
\caption{\label{fig:sputt_delta} (a) Evolution of cumulative number of sputtered atoms and total displacement in $z$-direction $\delta_z$ (in the inset) with number of ion impacts for the incident angles of \ang{0}, \ang{45}, \ang{70} and \ang{85}. (b) Evolution of total displacement in $x$-direction, $\delta_x$, with number of ion impacts for the same incident angles. Inset shows the evolution of the negative component $\delta_{x}^{-}$ up to 54\,000 impacts. The higher number of impacts for the grazing incidence of \ang{85} resulted in monotonous growth of this value with the same slope. Grey dashed lines guide the eye to show the deviation from a linear growth of the number of sputtered atoms and cumulative $\delta_{x}$ at \ang{45} and \ang{70} irradiation.}
\end{center}
\end{figure}


Figure \ref{fig:sputt_delta}(a) shows the evolution of the cumulative number of sputtered atoms with fluence for all the simulated cases, including the \ang{70} incidence. As expected, we see the highest sputtering dynamics under the ion beam tilted at \ang{70} with respect to the surface normal. This value grows linearly with the number of incident ions, which is the same as for the \ang{45} tilt. A slight increase of the growth rate is noted when the number of incident ions is larger than 40\,000 (a thin dashed line guides the eye in Figure \ref{fig:sputt_delta}(a)), when the surface develops significant roughening. Surprisingly, no such change is observed for the ions impacting on the surface at \ang{45}, although the roughening also appears at approximately the same fluence 
as clearly seen in Figure \ref{fig:pos_z_0d_45d}(c-d). The sputtering, and hence the erosive component of a surface self-organization in other two cases, is negligibly small ($\sim$ a few tens of atoms after 160000 Ar$^+$ impacts). Overall, the effect of sputtering appears negligible under the current irradiation condition (the highest sputtering yield per incoming ion is $\sim$ 0.03). Hence in the following we focus mainly on the effect of atomic displacements.

We analyze the total displacement calculated according to Equation (\ref{eq:delta}) along the $z$- and $x$-axes in the inset of Figure \ref{fig:sputt_delta}(a) and in Figure \ref{fig:sputt_delta}(b), respectively. The $y$ component of the total displacement is not included, since there is no dynamics in this direction and the value fluctuates around zero. The dynamics of evolution of $\delta_z$ with the fluence in all cases show rapid initial growth with the clear tendency towards saturation. The highest value of $\delta_z$ we observe for the normal incidence, where the saturation is also reached at much higher fluences than for any other incident angles. Strictly speaking, in our simulations we have not yet reached the saturation plateau, only significant slowing of the growth rate. The saturation for \ang{45} and \ang{70} incident angles is more apparent, although some growth of these value may still be expected in these cases as well. For the grazing incidence of \ang{85}, the growth still continues, even after a very high fluence. Under this irradiation condition, the displacements are rare and involve mainly individual atoms only. We note that the evolution of this component, although fairly distinct, does not show any correlation with surface self-organization, since all four plots have very similar behavior apart from the growth rate which is rising with increase of the $z$ component of the incident momentum. However, we can clearly correlate the dynamics of the growth of this parameter with increased roughness of the surface. This process is also important as it indicates the atomic flow above the initial surface level that later on can be displaced along the surface, if a driving force for this process is provided.

More interesting behavior of the total displacement in the system is seen in the evolution of $\delta_x$ component, which is shown in Figure \ref{fig:sputt_delta}{(b)}. Here, we see similarly fast growth of $\delta_x$ with the fluence for both \ang{45} and \ang{70} incident angles, although it is clear that the growth of $\delta_x$ slows down for the \ang{70} incidence, while it continues linearly increasing with the same initial rate for the ions impacting at \ang{45} incidence (consistent with the single-impact results shown in Ref. \cite{Lop18}). We draw the thin dashed lines to guide the eye along the linear increase of $\delta_x$ for both \ang{45} and \ang{70} cases. For the \ang{70} incidence, we observe a clear deviation from the linear behavior, which is explained by the accumulation of the negative displacement $\delta_x^-$ in backward direction, as it was discussed in Ref. \cite{Lop18a}, see the black line in the inset of Figure \ref{fig:histo_delta_atomic}(b). Here we note that similar deviation is also observed in case of the \ang{45} incidence, but the non-linearity in this case is much less pronounced and the ripple pattern does not emerge on the surface (see the red line in the inset of Figure \ref{fig:histo_delta_atomic}(b) and image in Figure \ref{fig:pos_z_0d_45d}(d)).  

Accumulation of $\delta_x$ under grazing incidence still shows some tendency of forward movement, however, the increase is insignificant. This result shows preferential transfer of momentum in the direction of the ion beam, although this bias is very small. $\delta_x$ for the normal incidence irradiation is omitted, since it was expected to fluctuate around zero.  

We note that the high value of $\delta_x$ found for the \ang{45} incidence does not convert directly to ripple formation (see Figure \ref{fig:pos_z_0d_45d} (c-d)). The single ion-irradiation model in Ref. \cite{Lop18}, on the other hand, predicts $\theta_{c}$=\ang{45} as a critical angle for ripple formation, while experimentally it is determined to be $\theta_{c}$ = \ang{55}. Our simulations are in line with experimental observations, since we clearly see that the strong total displacement $\delta_x$ under the \ang{45} of incidence is insufficient for ripple formation, i.e. the developed roughness does not align in this case into well-organized merged structures. 

Regarding the results shown in Figure \ref{fig:sputt_delta} {(b)} on \ang{85}, we notice that the contribution in $x$-direction fluctuates around zero. We can see that neither the accumulation of $\delta_x$ nor the sputtered atoms can directly explain the formation of the ripples over the surface, and, neither, the orientation of these. We have observed that the process starts with small displacements of the atoms, piling up and, after that, these structures merge and build the ripples. The displacements in general are short over the surface ($x-y$ plane). However, the small displacements outward from the surface ($\delta_z$) seem to explain the formation of ridges (note that $\delta_z$ is constantly increasing in Figure \ref{fig:sputt_delta} (a) inset). Due to the grazing incidence, the incoming ions induce shallow effects in the surface, but the high fluence enable to create these small structures aligned to the parallel direction (perpendicular mode). The height of these structures is low, as was observed experimentally \cite{Lop18}. 

\subsection{Displacements of surface atoms under high fluence irradiation}
\label{sec:discussion_toplayer}

Since in the previous section we saw that atomic displacements play more significant role in surface self-organization under the low energy ion irradiation, at least, in the following we focus on the analysis of the displacements of the surface atoms only. By surface atoms we understand the atoms which are found within the same layer as defined in Ref. \cite{Lop18a}, i.e. the atoms with $z$-coordinate greater than 22 \AA~. 
This layer includes all atoms that were actively displaced due to ion impacts at the surface, contributing to the mounds of the surface roughness. 

\begin{figure}[H]
\begin{center} 
\includegraphics[width=0.75\linewidth]{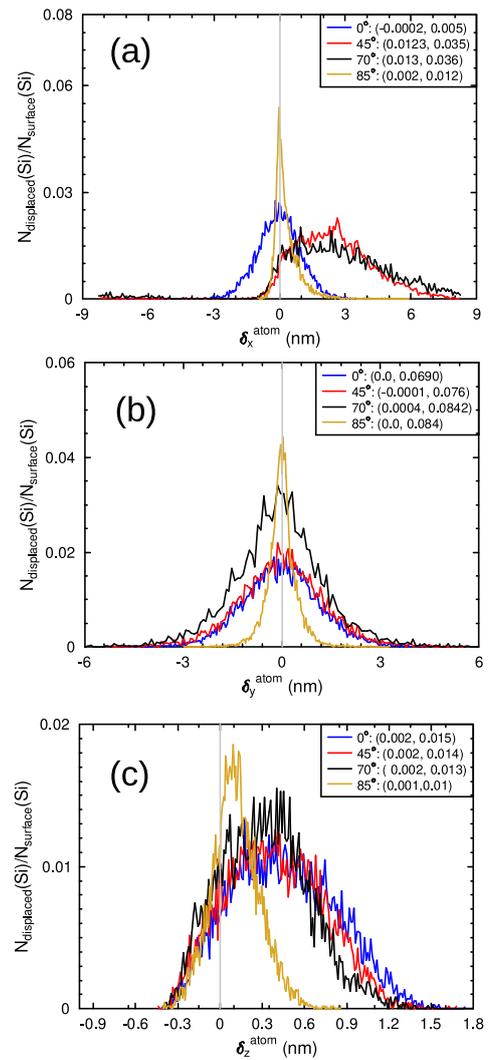}

\end{center}
\caption{\label{fig:histo_delta_atomic} Histograms of the ratio of displaced atoms ($N_{displaced} (Si)$) over the total number of atoms considered in the upper-most 22 Å of the cell ($N_{surface} (Si)$), sorted according to the atomic displacements {($\delta_{i}^{atom}=i_{final}^{atom}-i_{initial}^{atom}$)} in {(a) $i$=x, (b) $i$=y and (c) $i$=z directions at the end of the simulations. 
The gray thin line marks the zero displacement.} {The mean and standard deviation $(\mu,\sigma)$ of the distributions are shown in the legends.}}
\end{figure}

Figures \ref{fig:histo_delta_atomic}(a-c) show the histograms of the individual atomic displacements $\delta_{i}^{\text{atom}}$ ($i$ stands for $x$, $y$ or $z$) in the corresponding directions for the surface atoms that are found in the final configurations of the simulations with respect to the initial positions. 

We see that the distributions of $\delta_{x}^{\text{atom}}$ (Figure \ref{fig:histo_delta_atomic}(a)) for the inclined incidence with the \ang{45} and \ang{70} tilts are very similar with slightly greater number of atoms displaced to larger distances for the \ang{70} incidence. The broad distributions of $\delta_{x}^{\text{atom}}$ reveal stronger tendency for collective movement of atoms, for instance, when a formed random mound has been moved along the surface. The smaller the mound, the further the displaced atoms can move, explaining a slow decreasing slope for the larger displacement values. The process is observed for both the \ang{45} and \ang{70} incidences, however, the self-organization is stronger in case of the \ang{70} incidence, which explains the slight difference in behavior of both distributions for positive displacements. We see here that the $\delta_{x}^{\text{atom}}$ distribution for the \ang{45} incidence is more compact with a sharp peak at about 3 nm, which indicates that the mounds indeed could have been moving under this incidence, but only to relatively short distances. The displacements greater than 5 nm are less frequent than for the \ang{70} incidence, which allows for stronger separation between the individual ripples and, hence, clearer self-organization structure.

The normal incidence also resulted in rather broad peak centered around zero, while the peak for the \ang{85} incident angle is very narrow (Figure \ref{fig:histo_delta_atomic}(a)). The broad peak at the normal incidence is also explained by flow of atoms and piling up into mounds, but since there is no preferential momentum transfer along the surface, there is no preferential movement of these mounds and, hence, the atoms are displaced in both positive and negative $x$ directions with equal probability. The peak of $\delta_{x}^{\text{atom}}$ for \ang{85} is very sharp and also centered near zero similarly to the distribution at the normal incidence, however, we observe a clear skewness towards positive $\delta_{x}^{\text{atom}}$. This indicates that the displacements in the $x$ direction, under the grazing incidence, contributes to atomic flow forward to much lesser extent compared to lower ion incidence tilts, since the displacements are produced mainly in single impacts and direct momentum transfer to the collided atom with only insignificant effect of atomic flow along the surface in $x$-direction.

The distributions of atomic displacements in $y$-direction, $\delta_{y}^{\text{atom}}$, shown in Figure \ref{fig:histo_delta_atomic}(b), are symmetric around 0 as expected, since there is no preference in the displacement of atoms in this direction. Surprisingly, while the distributions collected for the atoms displaced under normal and \ang{45} incidence are almost identical, the number of displaced atoms side-wise at \ang{70} incidence is noticeably greater. This is an important observation for explaining the nature of self-organization and pattern formation, since it indicates the greater momentum transfer to the surface atoms causing setting in motion many atoms collectively in every ion impact. $\delta_{y}^{\text{atom}}$ for \ang{85} incidence is again distributed much narrower around zero as compared to the other distributions.

In the graphs of Figure \ref{fig:histo_delta_atomic}(c) we observe asymmetry in all distributions, which is explained by preferential relaxation of energetic atoms towards the open surface as well as formation of surface roughness via a pileup effect. However, some of the atoms where displaced inwards into the bulk, following the momentum transfer from the incoming ion beam. Remarkably, all distributions of $\delta_{z}^{\text{atom}}$ for non-grazing incidence (\ang{0} and \ang{45}) are very similar and extend to the same height. However, the distribution of $\delta_{z}^{\text{atom}}$ for \ang{70} incidence is slightly different, it shows more coherent displacements of atoms to the same distance, which is explained by formation of regular structures, such as ripples. Moreover, the ions incoming at \ang{70} to the surface normal were apparently able to displace the atoms forward in $x$ direction, but both in positive and negative $z$ directions. This specific way of atomic displacements provided stronger force for self-organization as the atoms under the surface were bound stronger and, hence, involved larger number of atoms in the collective movement. The grazing incidence resulted in the narrower distribution of $\delta_{z}^{\text{atom}}$, however, the shape and the bias towards the positive directions (outwards from the surface) is the same.  

\section{Discussion}
\label{sec:discussion}

Results presented in Section \ref{sec:results_analysis} clearly indicate that erosive component of the ion irradiation process plays insignificant role in surface modification processes in the considered ion energy regime. However, the dynamics of atomic displacements is very interesting and a few important observations can be deduced from its analysis. First of all, the displacements in $z$-direction keep increasing with fluence, although this increase slows down dramatically after surface roughness has developed. Organization of random surface features (mounds) into larger formations (ridges) slows down the growth of $\delta_z$ further. We see that the strongest decrease of the growth slope amongst the tilted incidence cases is observed for the inclination of \ang{70}. The grazing incidence resulted in even slower growth of $\delta_z$, which is, however, explained by much slower dynamics of modification of surface morphology compared to other studied cases. Analysis of the total displacement in the $y$-direction, $\delta_y$, did not reveal any interesting features.

The situation changes when the total displacement $\delta_x$ is analyzed. Here we see different behavior of increase of the cumulative total displacement with fluence. Although the inclination of \ang{70} results in the greater momentum projection on the $x$-axis, the growth of $\delta_x$ for \ang{70} is slower than that for \ang{45}. We explained this counter-intuitive result by analyzing the behavior of the negative component of $\delta_x^{-}$ for both angles. In the inset of Figure \ref{fig:sputt_delta}(b) we compare the evolution of $\delta_x^{-}$ under the \ang{45} and \ang{70} incident angles (the red and black curves, respectively). We see that initially, the growth of the cumulative backward displacement is equally rapid in both cases, which is explained by linear summation of all backward displacements in the non-overlapping individual cascades. This growth slows down again in both cases, however, more efficiently for the case of \ang{70}. This indicates the start of cascades overlaps, which is responsible for non-linear growth in cumulative displacements, discussed earlier in Ref. \cite{Lop18a}. Here we emphasize that the nonlinear effect in evolution of the total displacement of surface atoms under tilted ion beam increases with increase of the tilt, i.e. with increase of the momentum transfer component along the surface. 

Eventually, both $\delta_x^{-}$ saturate at different levels, since the shorter forward displacements that surface atoms experienced under the \ang{45} tilt allowed for accumulation of greater $\delta_x^{-}$ before the saturation. After $\sim$ 20\,000 ion impacts, $\delta_x^{-}$ starts growing again for the \ang{70} case with the same slope as for the \ang{45} case, but after $\sim$ 45\,000 ions (twice as many){, $\delta_x^{-}$ decreases again}. This indicates the start of self-organization, in other words, the formation of ridges from the random mounds. 
We also see that approximately at the same fluence (45\,000 ion impacts) as the one that induced the growth of $\delta_x^{-}$ under \ang{45} incidence, the slope of $\delta_x^{-}$ changes to much stronger one under the \ang{70} incident angle. The change of the slope to a more rapid one coincides with the start of a clear organization of the ripples on the surface {(for a more detailed process of ripples formation at \ang{70} incidence compared to \ang{45} and \ang{85}, please see Supplementary material)}. The large ridges protect backward displacements at the back of the ridge, while the forward displacements before the ridge are pushing the atoms forward. The random ridges on the surface formed under the \ang{45} incidence do not shield sufficiently the backward displacements (see inset of Figure \ref{fig:sputt_delta}(b)) and self-organization does not occur (see Figure \ref{fig:pos_z_0d_45d} (d)). 

Analysis of the atomic displacements $\delta_z^{\text{atom}}$ reveals that the flow of atoms in the $z$-direction, although fairly strong, does not explain self-organization of the surface into a pattern. We see that the distributions of $\delta_z^{\text{atom}}$ for the surface atoms in all studied cases are very similar, while the atomic displacements in this direction on the surface with the most organized structures are even slightly shorter than those observed on the surfaces with random mounds, see Figure \ref{fig:histo_delta_atomic}(c). Clearly the atomic flow in the $xy$ plane contributed to the self-organization to much greater extent. Although the total displacement $\delta_y$ fluctuates around zero and seemingly does not evolve with fluence, the distribution of atomic displacements $\delta_y^{\text{atom}}$ clearly indicates that the flow of atoms in this direction occurred as well (see Figure \ref{fig:histo_delta_atomic} (b)). Moreover, this flow is the strongest under the \ang{70} incidence, the case which resulted in the clear ripples on the surface. 

\begin{figure}[H]
\begin{center} 
\includegraphics[width=1\linewidth]{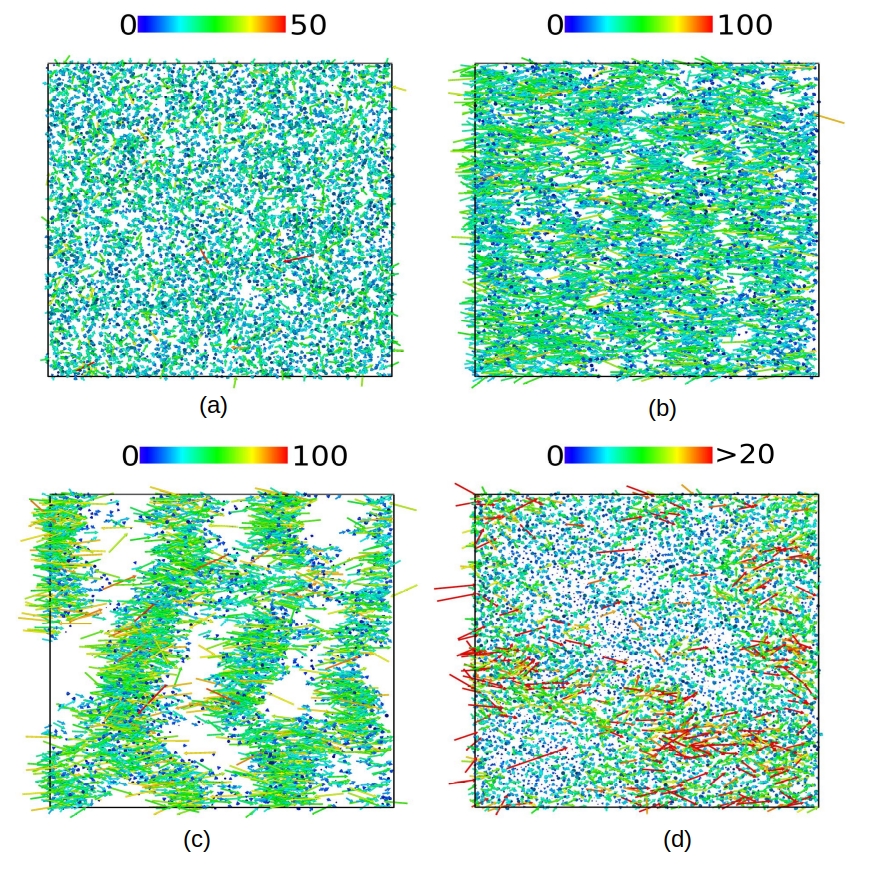}
\end{center}
\caption{\label{fig:top_surface_conf} Displacements evolution of the upper-most 22 Å of the cell from initial to final configuration for (a) 0$^{\circ}$ (54000 Ar$^{+}$) ($\alpha=45.4320025\pm0.434508$), (b) \ang{45} (54000 Ar$^{+}$) ($\alpha=22.3\pm0.2$), (c) \ang{70} \cite{Lop18a} (54000 Ar$^{+}$) ($\alpha=21.3\pm0.3$) and (d) \ang{85} (160200 Ar$^{+}$) ($\alpha=36.2\pm0.4$), colored according to their lengths {in \AA. The vectors are scaled 1/5 in (a-c) and 1/2.5 in (d) for visibility purposes.}} 
\end{figure}

To visualize the effect of atomic flow, in Figure \ref{fig:top_surface_conf} we show the atomic displacements as the vectors connecting the initial positions of the surface atoms and their positions in the final snapshots. The atoms shown in these Figures are the surface atoms with the $z$-coordinate above 22 \AA. The absolute value of these displacements can be estimated from the corresponding color bars. In these Figures we observe fairly small random displacement of atoms in the $xy$ plane under the normal incidence, see Figure \ref{fig:top_surface_conf}(a), with the more significant flow in the $z$-direction (note the color scale, but the small length of the vector projections on the $xy$ plane.) More intriguing situation can be seen from the comparison of Figures \ref{fig:top_surface_conf}(b) and \ref{fig:top_surface_conf}(c). Here the displacement vectors for the \ang{45} and \ang{70} incident angles are shown, respectively. Although the absolute values of these displacements are fairly close (displacements appear mainly in green color on the same color scale), we see that there are more vectors of yellow and red colors on the surface with the \ang{70} of the incidence, indicating stronger displacements of the surface atoms, i.e. ion-beam driven atomic flow of surface atoms. Clearly, higher mobility of surface atoms under the \ang{70} incidence contributed to self-organization of the ripples, while the same ripples under the \ang{45} of incidence are almost indistinguishable.

The distribution of atomic displacements under the \ang{85} incidence also reveal correlation between the position of the ridges formed parallel to the ion beam, and the concentration of the longer displacements than those of the atoms that remained in the trenches between the ridges (see Figure \ref{fig:top_surface_conf}(d)). Certainly, these atoms experienced very little displacements, while the atoms which ended up in the ridges travelled much longer distances mainly ahead but also strongly to the side. It has been suggested \cite{hofsass2014vel} that the transition of ripple mode from parallel to perpendicular under the grazing incidence occurs because of dominating role of erosion over mass redistribution in  the ripple formation process. This conclusion was based on analysis of low-to-medium range of irradiation energies (down to hundreds of eV), known to cause fairly strong sputtering of surface atoms under grazing incidence. In the present case, under the ultra-low irradiation energies, this argument does not hold as the sputtering yield is extremely low and surface erosion is very inefficient to explain surface patterning, which was observed also in experiment \cite{chowdhury2016tuning}.  In our simulations, we clearly see that the self-organization is still mainly a result of redistribution of surface atoms that becomes stronger the longer the grazing incidence irradiation is applied, see Figure \ref{fig:top_surface_conf} (d) and Supplementary movies \textsc{85degrees-surf-formation-disp.mp4} and \textsc{85degrees-surf-formation-posz.mp4}.

Furthermore, we estimated the deviation of the displacement direction from the ion beam direction, i.e. from the $x$-direction by comparing the $x$ and $y$ components of the displacements ($\alpha = tan^{-1}(\frac{\delta_y^{atom}}{\delta_x^{atom}})$). These values are given in the parentheses in the caption of Figure \ref{fig:top_surface_conf}. Surprisingly, we see a preferential deviation of the displacements with respect to the ion beam direction, which is similar for all atoms (the obtained error bar is given by the standard error of the mean). This shows that the component perpendicular to the direction of the beam plays an important role in the self-organization process. In both cases of \ang{45} and \ang{70} this deviation was about \ang{22}, which indicates preferential atomic movement in the $x$-direction, while for \ang{0} and \ang{85} these deviations are \ang{45} and \ang{36}, respectively. The former value shows no preference in the displacements in $x$- and $y$-directions, while the latter shows the stronger bias of displacements in the $y$-direction in self-organization process than that for the less tilted incidences.

\begin{figure*}[ht]
\centering
\includegraphics[width=0.9\linewidth]{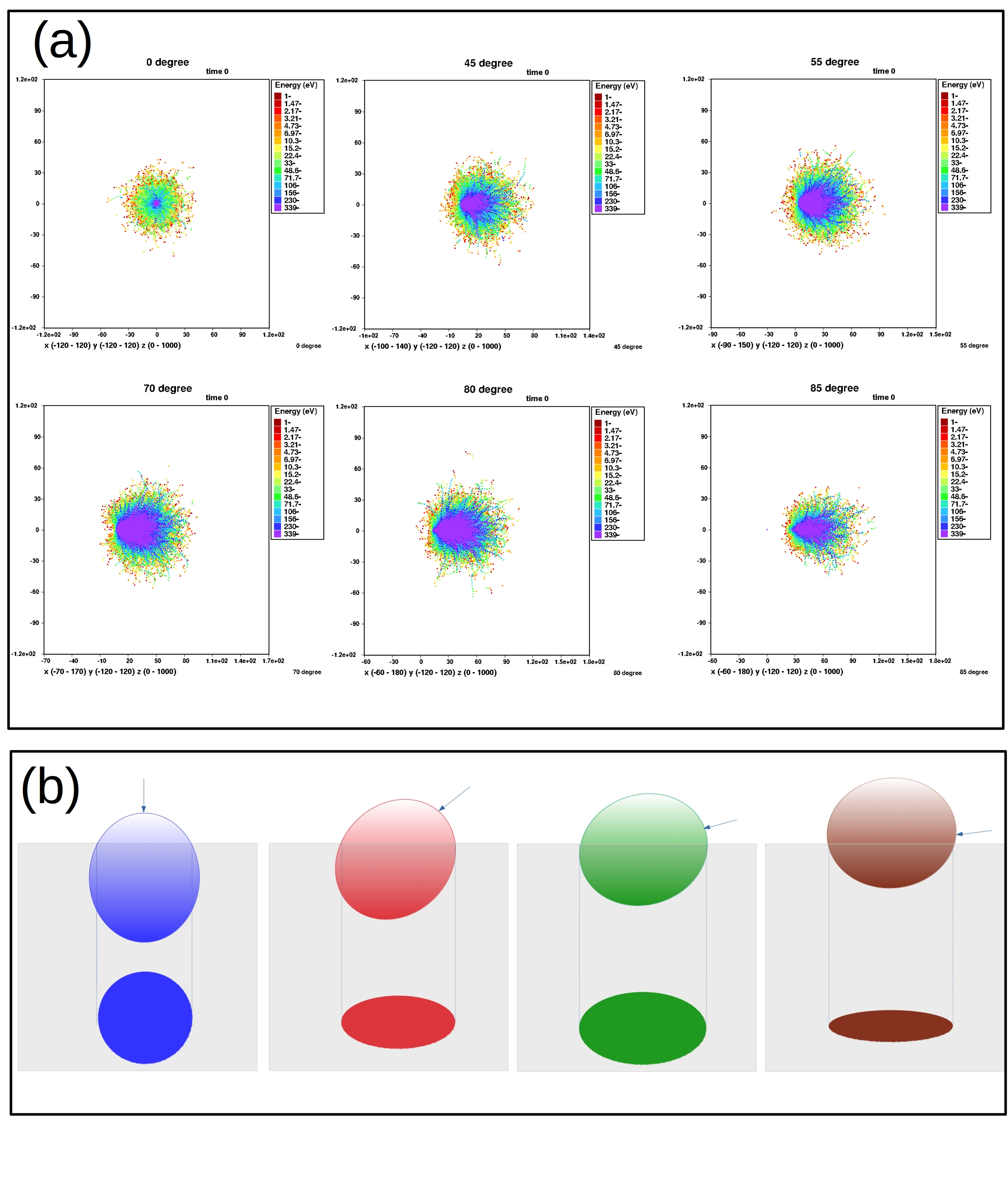}
\caption{\label{fig:bca_cascades} (a) Maps of trajectories of atoms in BCA cascades induced by 1 keV Ar on Si for \ang{0}, \ang{45}, \ang{55}, \ang{70}, \ang{80} and \ang{85}, from left to right and from top down. The trajectories are recorded only within the 5 \AA~ surface layer and include the recoils with energies above 1 eV. The trajectories were generated from 2000 statistical ion impacts. (b) Illustration of intersections of collision cascade regions, presented in the shape of an ellipse, with the surface for a given energy of incoming ions under \ang{0}, \ang{45}, \ang{70} and \ang{85} of incidence. The darker tone of the corresponding color represents the actual cascade inside the material (grey squares) and the lighter tone represents the missing part of the cascade. The ellipses below are the top views of the cross-sectional areas of these intersections.}
\end{figure*}

To clarify further the effect of atomic displacements, we performed the BCA calculations of the cascades for several incident angles. We used the CASWIN code \cite{bukonte2013}, which allows to trace the atomic trajectories observing statistical evolution of cascades. Since BCA is not reliable for energies below 1 keV~\cite{Lop17}, we have simulated the trajectories for 1 keV Ar ions in Si for all energetic atoms that received momenta in collision cascades within the very top surface layer of 5 \AA. Figure \ref{fig:bca_cascades} (a) summarizes these calculations showing the color-coded trajectories, where the color indicates the energy of the atom which it had at the given point of the trajectory. 

In these figures we can clearly see the size increase of the high energy spot on the two-dimensional map (see violet color in the middle of the cascades) with increase of the ion beam tilt. At the grazing incidence the spot decreases again and shrinks also laterally. The shape of the cascades under \ang{85} is the most elongated in the $x$-direction amongst all the presented cases. Moreover, we see that with increase of incidence the lateral spread of the cascade ($y$-direction) becomes broader (see also Figure \ref{fig:histo_delta_atomic} (b)), which can explain the more efficient self-organization as the atoms receive momentum not only along the $x$ axis, but also side-wise, with significant $y$ component. This observation is well in line with our MD results, where we see that both \ang{45} and \ang{70} incidences induce sufficient migration of atoms on surface, although the momentum transferred to surface atoms under the \ang{45} of incidence is not sufficiently strong to displace atoms to large distances. The shape of the cascades at \ang{85} is small and narrow indicating that  the cascades under this incidence are shallow. In these, the side-wise spread becomes dominant although the momentum transfer in $x$-direction exists. This explains the formation of perpendicular-mode of the ripples. 

In Figure \ref{fig:bca_cascades} (b) we illustrate schematically the change of the cascade shape within the surface with increase of the incident angle. Here the energetic part of a cascade is illustrated as an ellipsoid with the major axis aligned with the direction of the incoming ion. Below each ellipsoid, the top views on the cross-sectional areas of cascades at their intersection with the surface are shown. We see that the shape of these areas can be correlated with the momentum transfer along the surface both in $x$- and $y$-directions. Clearly the surface cross section of the collision cascade at \ang{70} has the largest area, hence the momentum transfer along the surface in both $x$- and $y$-directions is the strongest. The shallow cascade under the \ang{85} incidence results in the most narrow cross section at the surface explaining the tendency of digging the trenches parallel to the ion beam.

\section{Conclusions}

In conclusion, using molecular dynamics simulations we showed that the direct formation of ripples on the surface of amorphous silicon occurs only when the incident angle is greater than a threshold. For the first time we show rotation of ripple orientation with respect to the ion beam direction under the grazing incidence. We explain the dynamics of modification of surface morphology by analyzing the displacements of atoms in terms of total displacements and distributions of individual atom displacements. We clearly observe that roughness of the surface builds up as a result of atomic flow in collision cascades, however, for self-organization of the random surface features, significant atomic flow driven by tilted ion beams along the irradiated solid surface is needed. Moreover, the forward flow is complemented by the sidewise flow which is important for formation of ridges aligned at a \ang{90} angle to the ion beam direction. At grazing incidence the sidewise flow becomes stronger, but involves less atoms, while majority of atoms receive the momentum inwards that causes the effective "digging-in" effect forming trenches, while ridges aligned in the beam direction are formed by atoms displaced stronger on the surface. All processes require high fluences that result in displacement of large amount of atoms.

\section*{Acknowledgements}

The work was performed within the Finnish Centre of Excellence in Computational Molecular Science (CMS), financed by The Academy of Finland and University of Helsinki. Computational resources provided by CSC, the Finnish IT Center for Science as well as the Finnish Grid and Cloud Infrastructure (persistent identifier urn:nbn:fi:research-infras-2016072533) are gratefully acknowledged.

\end{document}